\PassOptionsToPackage{table,xcdraw}{xcolor}
\documentclass[10pt,conference]{IEEEtran}
\usepackage[table]{xcolor}  
\ifCLASSOPTIONcompsoc
  \usepackage[nocompress]{cite}
\else
  \usepackage{cite}
\fi
\ifCLASSINFOpdf
\else
\fi
\usepackage{amsmath}
\usepackage{balance}  
\usepackage{graphics} 
\usepackage{txfonts}
\usepackage{times}    
\usepackage[pdftex]{hyperref}
\usepackage{color}
\usepackage{textcomp}
\usepackage{booktabs}
\usepackage{ccicons}

\usepackage{cite}
\usepackage{url}
\usepackage{fancybox}
\usepackage{multirow}
\usepackage{flushend}
\usepackage{booktabs}
\usepackage{tabularx}
\usepackage{comment}
\usepackage{array}
\usepackage[flushleft]{threeparttable}
\usepackage{mdframed}
\graphicspath{{}{images/}{dia/}}
\DeclareGraphicsExtensions{.pdf,.png}

\usepackage{listings}
\usepackage{courier}
\usepackage{hyperref}

\usepackage{xpatch}

\xpatchcmd{\refstepcounter}{%
  \stepcounter{#1}%
}{%
  \stepcounter{#1}%
  \xdef\scr{\number\value{#1}}%
}{\typeout{success}}{\typeout{failure}}

\newcounter{o}
\usepackage{tikz}
\usepackage{styles/pgf-pie}
\usetikzlibrary{positioning,shadows}
\usepackage{balance}

\newif\ifpienumberinlegend
\pgfkeys{/number in legend/.code=
    \expandafter\let\expandafter\ifpienumberinlegend
    \csname if#1\endcsname
    \ifpienumberinlegend

    \def\beforenumber##1\afternumber{}%
    \fi,
    /number in legend/.default=true
}
\definecolor{1c1}{RGB}{188,162,6}
\definecolor{1c2}{RGB}{137,129,80}
\definecolor{1c3}{RGB}{239,167,31}
\definecolor{1c4}{RGB}{88,194,241}
\definecolor{1c5}{RGB}{6,180,188}

\tikzset{mynode/.style={draw=white,solid,circle,fill=green,inner sep=1pt, thick,
text=black}}
\tikzset{arrow line/.style={dashed, line width= 2.5pt, color=#1}}

\def\bf{\textbf}

\def\it{\textit}

\def\tt{\mct}

\newcommand{\mct}[1]{{\footnotesize {\texttt {#1}}}}

\usepackage{paralist}

\usepackage{caption}
\usepackage{subcaption}

\newcommand{\nd}{\vspace{1mm}\noindent}

\usepackage{tikz}

\makeatletter

\newcommand\notsotiny{\@setfontsize\notsotiny{6}{6}} 

\makeatother

\lstdefinestyle{inlinecode}{basicstyle={\ttfamily\scriptsize\bfseries}}

\newcommand{\urls}[1]{{\scriptsize\url{#1}}}
\usepackage{tcolorbox}

\usepackage{paralist}
\usepackage[outercaption]{sidecap}
\usepackage [autostyle, english = american]{csquotes}
\MakeOuterQuote{"}
\newcounter{scn}
\usepackage[shortlabels]{enumitem}
\usepackage{bchart}

\newcounter{finding_counter}
\newtoggle{comment}
\toggletrue{comment}

\usepackage{tcolorbox}
\newlist{RQ}{enumerate}{1}
\setlist[RQ, 1]{label = RQ \arabic*:}

\usepackage{forest}
\usepackage{soul}

\usepackage[multiple]{footmisc}
\usepackage{threeparttable} 

\usepackage{subcaption}
\usepackage{graphicx}

\begin{document}
\title{Combining Contexts from Multiple Sources for Documentation-Specific Code Example Generation}

\author{
\IEEEauthorblockN{Junaed Younus Khan and Gias Uddin\\DISA Lab, University of Calgary}
}













\IEEEtitleabstractindextext{%
\begin{abstract}
Code example is a crucial part of good documentation. It helps the developers to understand the documentation easily and use the corresponding code unit (e.g., method) properly. However, many official documentation still lacks (good) code example and it is one of the common documentation issues as found by several studies. Hence in this paper, we consider automatic code example generation for documentation, a direction less explored by the existing research. We employ Codex, a GPT-3 based model, pre-trained on both natural and programming languages to generate code examples from source code and documentation given as input. Our preliminary investigation on 40 scikit-learn methods reveals that this approach is able to generate good code examples where 72.5\% code examples were executed without error (passability) and 82.5\% properly dealt with the target method and documentation (relevance). We also find that incorporation of error logs (produced by the compiler while executing a failed code example) in the input further improves the passability from 72.5\% to 87.5\%. Thus, our investigation sets the base of documentation-specific code example generation and warrants in-depth future studies.

\end{abstract}

}

%


\maketitle

\IEEEdisplaynontitleabstractindextext
\section{Introduction}\label{sec:introduction}
Modern-day rapid software development is often facilitated by the reuse of software libraries/frameworks. The learning and usage of such libraries is supported by official documentation. While good documentation helps developers to use the corresponding library function, bad
documentation can severely harm their productivity~\cite{DeSouza-DocumentationEssentialForSoftwareMaintenance-SIGDOC2005, Robillard-APIsHardtoLearn-IEEESoftware2009a,Robillard-FieldStudyAPILearningObstacles-SpringerEmpirical2011a,Aghajani-SoftwareDocPractitioner-ICSE2020, Khan-DocSmell-SANER2021,Uddin-HowAPIDocumentationFails-IEEESW2015}. It is found that developers prefer documentation that contain code examples along with textual description \cite{Forward-RelevanceSoftwareDocumentationTools-DocEng2002, DeSouza-DocumentationEssentialForSoftwareMaintenance-SIGDOC2005}. Code examples help the developers understand the usage of the given code unit (e.g., method) with less effort and time \cite{Carroll-MinimalManual-JournalHCI1987a, Shull-InvestigatingReadingTechniquesForOOFramework-TSE2000}.

Despite developers' constant urge for code examples in documentation, many real-world documentation does not come with complementary code examples. In fact, the lack of code examples has been identified as one of the most severe issues in documentation \cite{Robillard-FieldStudyAPILearningObstacles-SpringerEmpirical2011a}. However, studies find that developers are often unwilling towards writing documentation as they find it less productive and less rewarding \cite{parnas2011precise, mcburney2017towards}, although developers do look for code examples when they seek to learn and reuse a software library~\cite{Uddin-HowAPIDocumentationFails-IEEESW2015}. Manual efforts for generating code examples is often labouring and time-consuming. Therefore, automatically generating code examples relevant to an API documentation unit might be a significant step towards improving documentation quality and usability. 

Several works have been done on automatic code generation based on given natural language (NL) intent (i.e., NL$\rightarrow$code). Such models are usually trained on input-output pairs of NL task description and corresponding code, or learn the mapping from naturally occurring corpora collected from different sources such as Stack Overflow, GitHub, etc. \cite{allamanis2015bimodal, yin2017syntactic, rabinovich2017abstract, xu2020incorporating, alon2020structural, feng2020codebert, wang2021codet5, austin2021program, parvez2021retrieval, nijkamp2022conversational}. However, none of these existing works has directly focused on generating code examples dedicated to complement documentation. Though somewhat related to Natural-language-to-Code synthesis, automatic generation of such documentation-specific code-examples is a comparatively new research direction. 

\begin{figure}[t]
  \centering
  \includegraphics[scale=.4]{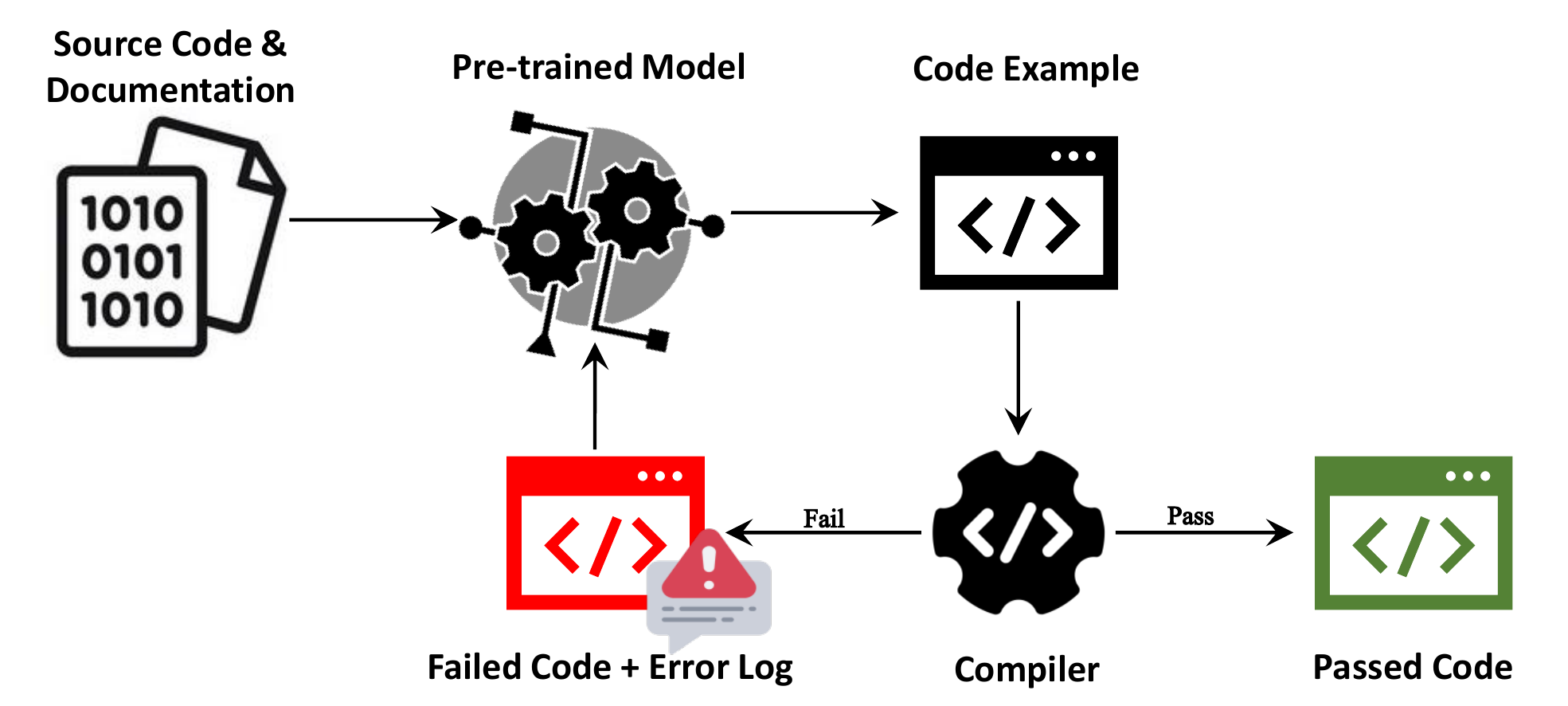}

  \caption{An overview of our approach}

  \label{fig:workflow_doc2code}
\vspace{-7mm}
\end{figure}
In this paper, we are focused on automatically generating code examples that can complement the documentation of a software library. Our goal is to include such code examples to the official documentation of a library. Hence, it is important that the produced code examples are correct, to the point, usable, and complete. It is also important that the code examples can fully address the problems specified in the documentation by taking into account the underlying development contexts. 
As such, our technique for code example generation uses contexts from different sources (such as source code, documentation, log contents) as input. Our proposed technique repeatedly uses the inputs to refine the underlying contexts, until a good quality code example is generated. An overview of our approach is shown in Figure \ref{fig:workflow_doc2code}. First, we pick a documentation that can have NL texts and the source code of a code unit (e.g., method). Second, we give this combination of source code and documentation to Machine Learning (ML) model that can produce a code example as a target. For the ML model, our initial experimentation has used the currently available pre-trained transformer model Codex \cite{chen2021evaluating}. Third, we attempt to compile the code example generated by the ML model. Fourth, if the code example generates error during the compilation, we instruct the ML model to fix and create a new code example by taking as input the error logs along with the failed code example itself. We repeat this process until the produced code example is compilable and relevant to the documentation contents. We answer two research questions to formally assess the quality of the produced code examples:

\noindent\textbf{RQ1. Can we generate good quality code examples for documentation using pre-trained models?} Transformer based pre-trained models are known for their state-of-the-art performances in several software engineering tasks such as automatic code synthesis, code repair, code summarization \cite{feng2020codebert,wang2021codet5, ahmad2021unified, phan2021cotext, khan2022automatic}. Hence, we feel motivated to explore transformer based model for the task at hand i.e., code example generation for documentation. We picked GPT-3 considering its huge success in different classification and generation tasks \cite{floridi2020gpt, chintagunta2021medically}, To be specific, we evaluated Codex which is a GPT-3 like model trained on over a dozen of programming languages like Python, Java, PHP, JavaScript, and so on \cite{chen2021evaluating}. We used the source code and the documentation of a method as input (prompt) of Codex and generated code examples in a zero-shot learning setting. Our intuition for the approach is simple and based on the naturalness of the pre-trained model. If a developer wants to write a code example for a documentation, s/he will first go through source code, understand the input/output format (i.e., parameters and return values). S/he will also refer to the corresponding documentation (if available) to get some further context. Hence, we design the input of Codex in a similar pattern.





\noindent\textbf{RQ2. Can we further improve the generated code examples using compiler generated log contents?}
Some of the generated code examples might not pass during execution because of compile time or run time error. In such cases, developers refer to compiler generated error messages in order to understand and fix their codes. Hence, we evaluate the ability of pre-trained model (Codex) to use such error messages to fix the error in the generated code examples. Using error logs as model's input for SE tasks is a new idea explored by some recent works \cite{zhang2022using, berabi2021tfix}. We provide the failed code examples and the corresponding error logs (produced by the compiler) as input of Codex and ask to fix them using prompt engineering.


While a full systematic evaluation of the proposed approach is our ultimate goal, we carry out this study as a preliminary investigation to get a heads-up about its feasibility and effectiveness. Hence, we conduct our study on 40 methods randomly selected methods from popular Python library scikit-learn. We evaluate the generated examples in terms of Passability (whether the generated example executes without error or not) and Relevance (whether the example properly deals with the targeted method and documentation). We find that Codex even with its' most basic setting (i.e., zero-shot learning) is capable of generating good quality code examples with 72.5\% passability and 82.5\% relevance. We further show that incorporation of error logs (produced by the compiler) improves the passability of the generated examples from 72.5\% to 87.5\%.

To the best of our knowledge, this is the first study that considers the task of documentation-specific code example generation. Though further investigation is necessary, this study proves the feasibility and effectiveness of our approach. We believe, it will help the research community to explore further in this direction.   

\textbf{Replication Package.} Our code and data are shared at \url{https://github.com/disa-lab/AutomaticCodeExample_SANER23}


\section{Related Work} \label{sec:related-work}


\subsection{Necessity of Code Examples in Documentation}
Studies suggest that example-based documentation is more beneficial to the developers than conventional Javadoc-style approaches~\cite{Cai-FrameworkDocumentation-PhDThesis2000,Carroll-MinimalManual-JournalHCI1987a,
Rossen-SmallTalkMinimalistInstruction-CHI1990a,Meij-AssessmentMinimalistApproachDocumentation-SIGDOC1992, DeSouza-DocumentationEssentialForSoftwareMaintenance-SIGDOC2005}. Shull et al. formed 15 teams out of 43 software enginnering students and asked them to perform some programming task in presence of both traditional and example-based documentation \cite{Shull-InvestigatingReadingTechniquesForOOFramework-TSE2000}. They found out that though the participants had complete knowledge about the usage of traditional documentation, they preferred the example-based ones as they were easier to understand. Forward and Lethbridge conducted a survey on software documentation involving 48 software practitioners which revealed that the text descriptions, examples, and their organization are the most important factors for documentation \cite{Forward-RelevanceSoftwareDocumentationTools-DocEng2002}. Nykaza et al. found that one main property of documentation contents desired by the software developers is the availability of easily understandable code examples \cite{Nykaza-ProgrammersNeedsAssessmentSDKDoc-SIGDOC2002}. Robillard and DeLine interviewed 440 software developers at Microsoft to identify the obstacles they face during learning new APIs \cite{Robillard-FieldStudyAPILearningObstacles-SpringerEmpirical2011a}. Their analysis showed that most issues were related to the official documentation such as the lack of code examples and the absence of task-oriented description.

\subsection{Automatic Code Generation}

Thanks to the amazing insight of Hindle et al. \cite{hindle2016naturalness}, language modeling of source code is now an emerging research area. Primarily, some n-gram based models have been proposed for code suggestion or auto-completion \cite{tu2014localness, nguyen2013statistical}. More recently, different NL$\rightarrow$Code generation models have been developed that generally work by training on bi-modal dataset containing NL and code pairs. Motivated by the bimodal models that map between images and natural language, Allamanis et al. considered similar type of mapping between natural language and source code snippets and built such a model leveraging NL and code pairs collected from Stack Overflow \cite{allamanis2015bimodal}. Yin and Neubig proposed a novel neural architecture which was accompanied by a grammar model to capture the underlying syntax of the target programming language \cite{yin2017syntactic}. Different transformer-based pre-trained models have also been employed for NL$\rightarrow$Code task such as CodeBERT \cite{feng2020codebert}, PLBART \cite{ahmad2021unified}, CoTexT \cite{phan2021cotext}. Research also shows that incorporating external knowledge collected from documentation, developers discussion, online forum, etc. can be useful to improve the performance of the code generation models \cite{hayati2018retrieval, xu2020incorporating,parvez2021retrieval, zhou2022docprompting}.   

Although several researches have been done on automatic code generation from a given natural language intent over the years, none of them directly focused on documentation-specific code examples. In this study, we consider this task and primarily evaluated our proposed approach.        




\section{Experiment Setup}\label{subsec:model_setup}






\textbf{Model.} OpenAI's GPT-3 is an auto-regressive language model with 175 billion parameters and 96 layers pre-trained on large-scale text data \cite{brown2020language}. Besides NLP, it has shown great promise in different downstream tasks of software engineering (SE). In fact, OpenAI has recently released a new descendant of GPT-3 dedicated for SE tasks named Codex \cite{chen2021evaluating}. Codex has been trained on both natural language and programming language (i.e., public code from GitHub). It has already been used to automate several SE tasks e.g., code generation \cite{finnie2022robots, trummer2022codexdb, xu2022systematic}, code summarization \cite{ahmed2022few, khan2022automatic}, code repair \cite{prenner2021automatic}, security bug-fix \cite{pearce2021can}. Therefore, we picked Codex for this study.    


\textbf{Prompt Engineering.} GPT-3/Codex works through prompt engineering, where the description of the task is embedded in the input (prompt) and the model (GPT-3) is expected to perform the task (e.g., text/code generation) accordingly. Prompting of GPT-3 can take place in different ways: \textit{zero-shot, one-shot, few-shot learning}. \textit{Zero-shot learning} calls for the model to produce a response without showing any examples. That means the prompt just contains the task description; no other information is provided. In contrast, \textit{one} (or \textit{few)-shot learning} involves giving one (or more than one) example(s) in the prompt. Though \textit{one} or \textit{few-shot learning} might yield better result, we have only experimented with \textit{zero-shot learning} in this exploratory study.

\textbf{Parameter Settings.}
GPT-3 involves several parameters. \textit{Temperature} and \textit{top-p} are two such parameters that control the randomness of the model's output ranging from 0 to 1. We set the temperature at a low value (0.2) while keeping top-P at 0.95 as used by Xu et al. \cite{xu2022systematic}. \href{https://beta.openai.com/docs/guides/code/best-practices}{Codex official documentation} also suggests similar setup. We keep \textit{Frequency} and \textit{Presence penalties} at their default values (0.0) that control the level of word repetition in the generated text by penalizing them based on their existing frequency and presence. In order to keep the code examples precise and simple, we select a moderately smaller \textit{max tokens size} of 256.

\section{Automatic Code Example Generation}
We present our observation based on our preliminary investigation of code example generation with pre-trained model (Codex) using source code, documentation, and error logs.

\subsection{RQ1. Contexts = Method Source Code + Documentation}
\subsubsection{Motivation}
A good code example is a vital part of documentation that helps developers to understand the usage of the underlying code unit (e.g., method) \cite{Nykaza-ProgrammersNeedsAssessmentSDKDoc-SIGDOC2002}. However, documentation of many software libraries/frameworks often lack suitable code examples which is marked to be a common documentation issue for the practitioners  \cite{Robillard-FieldStudyAPILearningObstacles-SpringerEmpirical2011a}. Hence, in this research question, we focus on documentation-specific code example generation using pre-trained model.   

\subsubsection{Approach}
While we could use any pre-trained model, we decided to evaluate Codex \cite{chen2021evaluating} for our preliminary study (as mentioned in Section \ref{subsec:model_setup}). We propose to use the source code and the documentation of a code unit (e.g., class, method) as the input (prompt) of Codex which will generate the code example accordingly. A sample input/output for our zero-shot learning is showed in Figure \ref{fig:code_example_generation} where we ask Codex to generate code example for the method {\ttfamily sklearn.metrics.cohen\_kappa\_score}. 

        
    
    

\begin{figure}
     \centering
     \begin{subfigure}[b]{\columnwidth}
         \centering
    \begin{lstlisting}[frame=single,keywordstyle=\color{black}]
`Method Source Code:`
    def cohen_kappa_score(y1, y2, *, labels, weights, sample_weight):
        confusion = confusion_matrix(y1, y2, labels, sample_weight)
        n_classes = confusion.shape[0]
        ... ... ... ... ... ... ... ... 
        ... ... ... ... ... ... ... ...
        ... ... ... ... ... ... ... ...
        k = np.sum(w_mat * confusion) / np.sum(w_mat * expected)
        return 1 - k
        
`Method Documentation:` 
    This function computes Cohen's kappa, a score that expresses the 
    level of agreement between two annotators on a classification 
    problem.
    
`Generate a code example for the above method and documentation:`
    ^[To be generated by Codex]^
    
    \end{lstlisting}
    
\caption{Input}
    \label{fig:zero_shot_code_example}
     \end{subfigure}
     \hfill
     
     \begin{subfigure}[b]{\columnwidth}
         \centering
        \begin{lstlisting}[frame=single]
# Import the cohen_kappa_score function 
`from sklearn.metrics import cohen_kappa_score`

# Define two annotators
`annotator_1 = [1, 1, 0, 0, 1, 1, 0, 0, 0, 0]`
`annotator_2 = [1, 1, 0, 0, 1, 0, 0, 0, 0, 0]`

# Calculate Cohen's kappa
`cohen_kappa = cohen_kappa_score(annotator_1, annotator_2)`

# Print the Cohen's kappa score
`print(cohen_kappa)`
    \end{lstlisting}
    
    \caption{Output}
    \label{fig:RQ1_example}
     \end{subfigure}
        \caption{Code example generation}
        \label{fig:code_example_generation}
        \vspace{-6mm}
\end{figure}    

While evaluation of large scale data should be the ideal goal, we experiment with a smaller number in this study to get a primary insight about the approach. To be specific, we focused on Scikit-learn (sklearn), a well-known and widely used Python library. We randomly selected 40 methods, collected their source codes and documentation, and generated code examples for them.

We plan to check three aspects of the generated code examples i.e., Passability, Relevance and Usefulness. For Passability, we run the generated code on compiler and note whether it passes (i.e., executes without error) or not. Relevance signifies whether the code example properly deals with the target method and documentation. Usefulness indicates the extent to which the generated example helps the developers. For this study, we manually analyzed the Passability and Relevance of the generated examples. We plan to analyze the Usefulness in the future. We also intend to design and employ metrics (e.g., BLEU score, Pass@\textit{k}) for automatic evaluation.

\subsubsection{Observation}
We find that passability of the generated examples is 72.5\% i.e., 29 out of 40 generated code examples were executed without any error. In terms of relevance, 82.5\% (33 out of 40) code examples successfully showed the usage of the target method. Figure \ref{fig:RQ1_example} depicts a code example for the method {\ttfamily sklearn.metrics.cohen\_kappa\_score} generated by Codex. This is fascinating how Codex understands the context of \textit{cohen's kappa coefficient} (used to measure agreement between two annotators \cite{mchugh2012interrater}) from the documentation and names the two lists as $annotator\_1$ and $annotator\_2$.








\subsection{RQ2. Contexts = Method Source Code + Doc + Error Logs}
\subsubsection{Motivation}
Whenever a code fails to execute due to compile/run time error, the developer looks at the error log first. Sometimes the developer immediately understands the mistake (based on her/his prior knowledge) and solves it, or s/he looks for a solution in online forums (e.g., Stack Overflow). Since transformer model like GPT-3/Codex is pre-trained on large amount of data collected from different sources (such as GitHUb, Stack Overflow), we anticipate that these models might be able to understand such error logs and fix the errors based on that. Hence, in this research question, we check if we can bring further improvement to the quality (i.e., passability) of the code examples (that failed in RQ1) using the error logs as input.

\subsubsection{Approach}
We plan to use the log contents (i.e., error messages) produced by the compiler to fix the code examples that were unable to pass in RQ1. We incorporate the error messages with the produced code examples in the GPT-3 prompt and tell it to fix the error as showed in Figure \ref{fig:code_example_improvement}. The goal here is to evaluate GPT-3's ability to fix a faulty code (given the error messages) and thus improve the Passability of the generated code examples. We do this for the 11 code examples that failed to execute in the generation phase (RQ1) and see how many of them can be fixed by this approach.


    
    
    

\begin{figure}
     \centering
     \begin{subfigure}[b]{\columnwidth}
         \centering
    \begin{lstlisting}[frame=single,keywordstyle=\color{black}]
`Code with error:`
    `from sklearn.pipeline import make_pipeline`
    `from sklearn.svm import SVC`
    `from sklearn.preprocessing import StandardScaler`
    
    # Create a pipeline object
    `pipe = make_pipeline(StandardScaler(), SVC(random_state=42))`
    
    # Fit the pipeline to the training data
    `pipe.fit(X_train, y_train)`
    
    # Score the accuracy on the test set
    `accuracy = pipe.score(X_test, y_test)`
    
    # Prints the accuracy
    `print('{0:.1%} test set accuracy'.format(accuracy))`

`Error Message:` 
     ~NameError: name 'X_train' is not defined~
    
`Correct the code based on the error message:` 
    ^[To be generated by Codex]^
    
    \end{lstlisting}
    
\caption{Input}
    \label{fig:input_log_prompt}
     \end{subfigure}
     \hfill
     
     \begin{subfigure}[b]{\columnwidth}
         \centering
        \begin{lstlisting}[frame=single]
`from sklearn.pipeline import make_pipeline`
`from sklearn.svm import SVC`
`from sklearn.preprocessing import StandardScaler`
`from sklearn.model_selection import train_test_split`
`import numpy as np`

# Create a pipeline object
`pipe = make_pipeline(StandardScaler(), SVC(random_state=42))`

# Create train and test sets
^X, y = np.arange(20).reshape((10, 2)), range(10)^
^X_train, X_test, y_train, y_test = train_test_split(X, y, test_size=0.3, random_state=42)^

# Fit the pipeline to the training data
`pipe.fit(X_train, y_train)`

# Score the accuracy on the test set
`accuracy = pipe.score(X_test, y_test)`

# Prints the accuracy
`print('{0:.1%} test set accuracy'.format(accuracy))`
    \end{lstlisting}
    
\caption{Output}
    \label{fig:output_log_prompt}
     \end{subfigure}
        \caption{Improvement using error message}
        \label{fig:code_example_improvement}
        \vspace{-6mm}
\end{figure} 

\subsubsection{Observation}
Incorporation of log contents (i.e., error messages) in the prompt immediately fixed more than half of the failed code examples (6 out of 11) from RQ1. Hence, the passability went up from 72.5\% to 87.5\%. We can see an example in Figure \ref{fig:code_example_improvement}. At first, the generated code example had an error since it was referring to some variables (such as $X\_train$) without defining them. After using the error message along with the failed code example as input (Figure \ref{fig:input_log_prompt}), the model now defines $X\_train$ and other variables that were previously undefined (highlighted with blue in Figure \ref{fig:output_log_prompt}). Thus the error got fixed.

\section{Conclusion and Future Work} \label{sec:conclusion}
In this paper, we focused on automatically generating documentation-specific code examples, a new domain in software engineering (SE) research. Initially, we investigated the feasibility and effectiveness of pre-trained model (Codex) for this task. We used source code and documentation as the input of Codex to generate code examples. Codex could generate code examples with 72.5\% passability and 82.5\% relevance with very basic setting of zero-shot learning. We also find that using error logs can be useful to further improve the passability of the examples that failed in the generation phase. Hence, this study sets up the foundation for future researches in this direction. In future, we intend to conduct more in-depth evaluation of our proposed approach. We plan our future studies from three different aspects- \textbf{1) Data:} Since the main goal of this study was to primarily verify our approach, we used a small set of data i.e., 40 Python methods and documentation. We need to study larger set of data coming from various sources and languages in the future. \textbf{2) Model:} \textit{M1.} We limited this study to zero-shot learning only. We will extend the investigation by also analyzing one/few-shot learning. \textit{M2.} Pre-trained transformer models (like GPT-3) also support fine-tuning. However, latest versions of GPT-3 models (e.g., Codex) are not available for fine-tuning. We could fine-tune GPT-3 when it is available to do so. \textit{M3.} We will investigate the effect of different parameters (associated with GPT-3) on the quality of generated examples which is not done in this study. \textit{M4.} We will leverage relevant external knowledge retrieved from online forums (such as Stack Overflow, GitHub) to improve the generated examples. \textbf{3) Evaluation:} \textit{E1.} We only used manual evaluation metrics (e.g., passability, relevance) in this study. We plan to use automatic metrics (like BLEU score, Pass@\textit{k}) in the future which would facilitate the evaluation of large scale data. However, for that we will also need a benchmark of input-output pairs (e.g., source-code, documentation$\rightarrow$code examples). \textit{E2.} We will conduct a user study (e.g., developers' survey) to evaluate the usefulness of the generated examples.

\nd\textbf{Acknowledgement.} This work was funded by Natural Sciences and Engineering Research Council of Canada, University of Calgary, Alberta Innovates, and Alberta Graduate Excellence Scholarship.

\begin{small}
\bibliographystyle{abbrv}
\bibliography{consolidated}
\end{small}

\end{document}